\documentclass{elsart}
\usepackage{times}
\usepackage{graphicx}
\usepackage{amsmath}
\usepackage{amssymb}
\newtheorem{theorem}{Theorem}

\newcommand{\quot}[1]{\lq#1\rq}

\begin{document}
\begin{frontmatter}
  
  \title{Variant Monte Carlo algorithm for driven elastic strings in random
    media} \author[ia1]{Alberto Rosso} \ead{rosso@physics.unige.ch} and
    \author[ia2]{Werner Krauth} 
  \address[ia1]{Universit\'e de Gen\`eve 4, DPMC, 24 Quai Ernest Ansermet,
    CH-1211 Gen\`eve 4, Switzerland} 
  \address[ia2]{CNRS-Laboratoire de
    Physique Statistique de l'Ecole Normale Sup{\'e}rieure,\\ 24 rue
    Lhomond, 75231 Paris Cedex 05, France} \ead{werner.krauth@ens.fr}

\begin{abstract}  
  We discuss the non-local \quot{Variant Monte Carlo} algorithm which has been
  successfully employed in the study of driven elastic strings in disordered
  media at the depinning threshold.  Here we prove two theorems,
  which establish that the algorithm satisfies the crucial
  \quot{no-passing} rule and that, after some initial time, the
  string exclusively moves forward.  The Variant Monte Carlo
  algorithm overcomes the shortcomings of local methods, as we show by analyzing
  the depinning threshold of a single--pin problem.
\end{abstract}
\end{frontmatter}

\section{Introduction}
\label{sec1} 

Over the past few years relevant progress has been made in the
study of disordered elastic interfaces.  One of the most intriguing
problems concerns the response of the interface to an external force
$f$. Two regimes are observed at zero temperature: (i) When $f$
is smaller than a certain critical threshold $f_c$, the interface
is pinned; (ii) when the force $f$ passes the threshold value
($f > f_c$) the system undergoes the so-called depinning transition
\cite{kardar}, which has been widely investigated during the
last years \cite{nattermann_stepanow_depinning,narayan_fisher_depinning}.
The functional renormalization group has allowed to gain a much deeper
understanding of this transition \cite{2loop}. A number of experiments
on the contact line of a liquid meniscus on a rough substrate
have also been analyzed \cite{robbins87,rosso_width2,moulinet_contact_line}.

In this context, we have introduced new algorithms which allow us to solve
the depinning problem at zero temperature in finite samples. This paper
discusses mathematical aspects of the Variant Monte Carlo (VMC) algorithm,
which have not been published yet, although they were implied in past
works \cite{rossokrauthI,rosso_krauth_non-harmonique,Goodman}.  In section
\ref{vmc}, we prove the no-passing theorem and the forward-moving property
for the VMC algorithm, then in section \ref{pin} we discuss in detail
the single--pin problem.

The VMC  algorithm is able to detect the critical force and the
critical configuration (i.e. the ultimate pinned configuration) of a
one-dimensional string with short-range elasticity moving on a disordered
two-dimensional lattice.  We have also developed a continuous algorithm
\cite{rosso_krauth_longrange}, which is remarkably useful in higher
dimensions and allows to handle long-range interactions.

\section{Variant Monte Carlo Algorithm}
\label{vmc}
 
We consider a string $h^t = \{ h_i^t\}_{i=0,\ldots,L}$ moving
at times $t=0,1,2,\ldots$ on a spatial square lattice of side $L$, in   
a random potential $V(i,h)$ with $h=0,\ldots,M$.  The energy
of the string is
\begin{equation}
E(h^t)=\sum_{i=1}^{L} \left[ V(i,h_i^t)-f h_i^t
+E_{\text{el}}(|h^t_{i+1}-h^t_i|) \right] , 
\label{energy} 
\end{equation}
where $f$ is the external driving force and 
$E_{\text{el}}$ a short-range convex elastic energy. We assume 
toroidal boundary conditions with a winding term for $f$ such that, 
at large $f$, the line keeps winding around the torus, lowering the 
energy at each time step.
In section \ref{pin}, we show that 
a non-local algorithm needs to be used allowing
an arbitrary number of points $i$ to move simultaneously by one site in any direction
\cite{rossokrauthI}.

Following Ref.~\cite{rossokrauthI} we define a \quot{forward front}
of length $k$ as a contiguous set of points $i,i+1,....,i+k-1$
which move together in forward direction.  A \quot{backward front}
is defined similarly.  A front is \quot{unstable} if moving it lowers
the energy (\ref{energy}).  One has to check only $ \sim 2L^2$ fronts
to establish whether a string is pinned, ie has no unstable front.
The \quot{depinning force} $f_d(h^{\alpha})$ of a string $h^{\alpha}$
is the smallest non-negative $f$ which destabilizes one of its forward
fronts. The \quot{critical force} of a whole sample can then be defined
as the largest of the depinning forces of all the strings in the sample
\begin{equation} f_c = \max_{ \{ h^{\alpha} \} }
  f_d(h^{\alpha}).
  \label{fcrit} 
\end{equation} 
The VMC algorithm simply moves a single front of \emph{minimal} length
$k$ among the unstable forward and backward fronts. The VMC algorithm is
not a valid Monte Carlo algorithm. However, each possible move within
the VMC algorithm is also allowed with all the non-local algorithms,
and its depinning threshold  and ultimate pinned configuration is the
same as the one of each non-local algorithm.  We will show that the above
definition of $f_c$ is appropriate for the VMC algorithm. This implies
that it is also correct for general non-local rules, even if they are
not restricted to moving fronts only (\cite{rossokrauthI}).

We stress that these definitions and the following
theorems can be easily extended to a general $d$-dimensional interface
with a convex elastic energy. However, the VMC seems to be practically
useful only for one-dimensional interfaces with
short-range elastic energy as the total number of fronts remains
polynomial. We prove the following theorems:
\begin{figure}
\centerline{\includegraphics[width=14 cm]{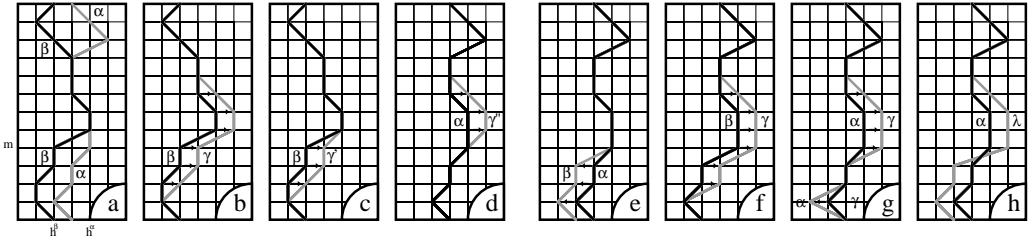} }
\caption{Illustration of the two theorems.}
\label{due}
\end{figure}
\begin{theorem}
  Let $h^{\alpha}$ be a stable configuration. Then the VMC algorithm
  cannot reach a configuration $h^{\gamma}$ with $h_i^{\gamma} >
  h_i^{\alpha}$ for some $i$ from a starting configuration  $h^{\beta}$
  with $ h^{\beta}_i \le h^{\alpha}_i \; \forall i$.
\label{theorem_one}
\end{theorem}
This theorem is illustrated in figure~\ref{due}($a$, $b$, $c$ and
$d$). We suppose the existence of a string $h^{\gamma}$
forbidden by the theorem (see figure \ref{due}$b$):
\begin{align}
  I)\ & E(h^{\gamma})-E(h^{\beta}) < 0 \nonumber \\
  II)\ & h^{\gamma}_{i} > h^{\alpha}_{i}\quad \text{for some $i$}.
\label{absurdeII}
\end{align}
Due to the definition of the VMC dynamics the front connecting $h^{\beta}$
to $h^{\gamma'}$ (see figure \ref{due}$c$) must be stable. Moreover, the
stability of $h^{\alpha}$ assures that the front connecting $h^{\alpha}$
to $h^{\gamma''}$ (see figure \ref{due}$d$) is also stable:
\begin{align}
  I)\ & E(h^{\gamma'})-E(h^{\beta}) > 0\nonumber  \\
  II)\ & E(h^{\gamma''})-E(h^{\alpha}) > 0 \,.
\label{dynamiqueconditionII}
\end{align}
Subtracting  Eq.~(\ref{absurdeII}{\it I}) from the equations
(\ref{dynamiqueconditionII}) and using the expression (\ref{energy}) leads to
\begin{equation}
E_{\text{el}}(|h^{\alpha}_m-1|)-E_{\text{el}}(|h^{\alpha}_m|)
+E_{\text{el}}(|h^{\beta}_m+1|) -E_{\text{el}}(|h^{\beta}_m|) >0 ;
\label{finalcondition}
\end{equation}
where, without any loss of generality, we set
$h^{\beta}_{m+1}=h^{\alpha}_{m+1}=0$ (see figure \ref{due}$a$).
From $h^{\beta}_m <h^{\alpha}_m$ we write:
\begin{align}
h^{\beta}_m \le  h^{\alpha}_m-1 <h^{\alpha}_m \nonumber \\
h^{\beta}_m < h^{\beta}_m+1 \le h^{\alpha}_m .
\label{finalconditionII}
\end{align}
A convex function $f(x)$ in $[x,y]$ satisfies for all $x_1,x_2 \in [x,y]$:
\begin{equation}
f(tx_1+(1-t)x_2)\le tf(x_1)+(1-t)f(x_2) \;\;\;\; 0 \le t\le 1.
\label{convexityI}
\end{equation}
Taking $x_1=h^{\beta}_m$ and $x_2=h^{\alpha}_m$, from (\ref{finalconditionII})
we can find a $t$ such that:
\begin{align}
 h^{\alpha}_m-1 = t \, h^{\beta}_m + (1-t) \, h^{\alpha}_m \nonumber \\
h^{\beta}_m+1 = t \, h^{\alpha}_m + (1-t) \, h^{\beta}_m .
\label{convexityII}
\end{align}
Using this relation to impose the convexity of the elastic energy in
(\ref{finalcondition}) we end up with a contradiction which demonstrates the
theorem.
\begin{theorem}
  Let $h^{\alpha}$ be pinned 
  in forward direction. Then, the VMC algorithm can at most recede towards a
  string $h^{\beta}$ ($h_i^{\beta} \le h_i^{\alpha} \; \forall i$),
  which is itself pinned in forward direction. The analogous property holds 
  for strings pinned in backward direction.
\label{theorem_two}
\end{theorem}
An illustration of this theorem is displayed in figure \ref{due}($e$, $f$,
$g$ and $h$).  We suppose that a configuration $h^{\gamma}$, forbidden by the
theorem, is reached by the string:
\begin{align}
  I) & E(h^{\gamma})-E(h^{\beta}) < 0 \nonumber \\
  II) & h^{\gamma}_{i} > h^{\alpha}_{i}\quad \text{for some $i$.}
\label{absurdeI}
\end{align}
The instability of the front connecting $h^{\alpha}$ to $h^{\beta}$ (figure
\ref{due}$e$) implies
\begin{equation}
 E(h^{\beta})-E(h^{\alpha}) < 0 \,.
\label{dynamiqueconditionI}
\end{equation}
We may connect the string $h^{\alpha}$ to the string $h^{\gamma}$ by
moving two fronts (see figure \ref{due}$g$) \footnote{In analogy with 
 theorem \ref{theorem_one} we have also to consider the case shown in figure
 \ref{due} $h$,
  where we suppose that the line starting from $h^{\beta}$ moves to 
  $h^{\lambda}$ instead of $h^{\gamma}$.
  Using the convexity condition (\ref{convexityI}) the theorem remains valid.}.
The forward front is stable because $h^{\alpha}$ is pinned in forward direction.
The backward front is also stable because it is smaller than the front
connecting $h^{\alpha}$ to $h^{\beta}$ (VMC dynamics).  Thus we conclude:
\begin{equation}
 E(h^{\gamma})-E(h^{\alpha}) > 0.
\label{VMCconditionI}
\end{equation}
Eq.~\ref{VMCconditionI} contradicts the sum of (\ref{absurdeI} {\it I}) and
(\ref{dynamiqueconditionI}), invalidating the starting hypothesis and proving
the theorem.

The \quot{no-passing theorem} (theorem \ref{theorem_one}) assures
that the VMC algorithm connects an arbitrary initial state with the
critical string, whereas the \quot{forward-moving theorem} (theorem
\ref{theorem_two}) allows us to understand that Eq.~(\ref{fcrit}) is
indeed appropriate: one might have imagined that the elastic line which
cannot advance at $f_c$ could move backwards and then be avoided during
the subsequent forward evolution. Theorem \ref{theorem_two} excludes
the existence of such loopholes.  Finally we remark that both theorems
transpose correctly to the lattice the analytical properties of the
continuum equation of motion \cite{middleton_theorem}.

\section{Single--pin problem}\label{pin}
\begin{figure}
\centerline{\includegraphics[width=10 cm]{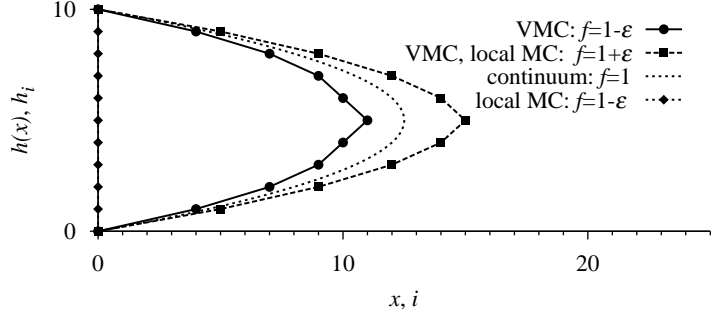} }
\caption{Single--pin problem with harmonic elasticity.
The stationary solution of the  VMC algorithm is close to the continuum
solution. The stationary solution of the local Monte Carlo algorithm 
approaches the continuum solution only for $f > f_{\text{loc}}=1$.  }
\label{single}
\end{figure}
We now discuss the motion both of a continuous and a discrete elastic
line. The line is pinned at a single point which corresponds, with
periodic boundary conditions, to $h_0=h_L=0$. The equation of motion
for the continuous line is
\begin{equation}
\partial h/ \partial t = - \partial E_{\text{el}}/\partial h + f.
\label{continuum}
\end{equation}
A discrete \emph{harmonic} elastic energy $E_{\text{el}}=
\frac{1}{2} | h_{i+1}-h_i|^2$ then corresponds to the continuous
energy $ E_{\text{el}}(x)=\frac{1}{2} (\partial h/ \partial x)^2$, to
be integrated over $x$.  The stationary solution under the indicated
pinning condition is easily seen to be $h(x)= \frac{1}{2} f x (L-x)$,
as shown in figure~\ref{single} for $L=10$.

We may also follow the dynamics of the discretized problem from a starting
configuration $h_i^{t=0}=0\; \forall i$. Of particular interest is the
case $f=1 \pm \epsilon$, with $\epsilon \gtrsim 0$. The VMC solutions
are given in the figure. They have to be contrasted with the solutions
of the local algorithm (only fronts of length $1$ can be moved). For
$f=1-\epsilon$, the starting configuration is stable under local dynamics,
as any forward move of a single point costs an elastic energy $1$, more
than what is recovered through the driving force.  The VMC solution is
recovered only for $f> f_{\text{loc}}1$ This specific \quot{critical
force} $f_{\text{loc}}$ of the local algorithm is independent of $L$
only for the harmonic elastic energy.

The local algorithm is more pathological for stronger than harmonic
elastic energies, which have proven to be important in this context
\cite{rosso_krauth_non-harmonique}.  This is evident in the metric
constraint model (see \cite{rossokrauthI}).  As an example (stronger than
harmonic, weaker than metric constraint) we treat a
quartic elastic energy $E_{\text{el}}= \frac{1}{12} | h_{i+1}-h_i|^4$
(corresponding to a continuous energy $ E_{\text{el}}(x)=\frac{1}{12}
(\partial h/ \partial x)^4$).  The stationary continuum
solution of Eq.\ref{continuum}: $h(x)= \frac{3}{8} (\frac{3 f}{2})^{1/3}
[ L^{4/3} - (L- 2x)^{4/3}]$, is again recovered with the
VMC algorithm. However, the specific critical force at which the
local algorithm becomes equivalent to the VMC method grows with $L$
as $f_{\text{loc}} \sim (\frac{3}{2} f)^{2/3} L^{2/3}$, \emph{i.e.}
diverges with $L$.

We conclude that the local dynamics is inconsistent even for a single-pin
problem. In general disordered samples, it similarly fails to describe
the real dynamics of the continuum and is very sensitive to exceptional
(local) configurations of the disorder potential, which can block
the string even of an infinite systems, and eliminate the interplay
between disorder and (collective) elasticity which is at the heart of
the depinning problem.


\end{document}